\documentclass[usenatbib,onecolumn]{mnras}
\usepackage{mathptmx,tabularx,paralist}

\DeclareRobustCommand{\VAN}[3]{#2}
\let\VANthebibliography\thebibliography
\def\thebibliography{\DeclareRobustCommand{\VAN}[3]{##3}\VANthebibliography}

\usepackage{graphicx,amsmath,amssymb,epsfig,comment}

\title{Resolved imaging of exoplanets with the solar gravitational lens}

\author[S. G. Turyshev and V. T. Toth]{
Slava G. Turyshev$^{1}$, Viktor T. Toth$^2$\\
$^1$Jet Propulsion Laboratory, California Institute of Technology,
4800 Oak Grove Drive, Pasadena, CA 91109-0899, USA\\
$^2$Ottawa, Ontario K1N 9H5, Canada}

\date{Accepted XXX. Received YYY; in original form ZZZ}

\pubyear{2022}

\begin{document}

\label{firstpage}
\pagerange{\pageref{firstpage}--\pageref{lastpage}}

\maketitle

\begin{abstract}
We discuss the feasibility of direct multipixel imaging of exoplanets with the solar gravitational lens (SGL) in the context of a realistic deep space mission.  For this, we consider an optical telescope, placed in the image plane that forms in the strong interference region of the SGL. We consider an Earth-like exoplanet located in our immediate stellar neighborhood and model its characteristics using our own Earth. We estimate photon fluxes from such a compact, extended, resolved exoplanet. This light appears in the form of an Einstein ring around the Sun, seen through the solar corona. The solar corona background contributes a significant amount of stochastic noise and represents the main noise source for observations utilizing the SGL. We estimate the magnitude of this noise. We compute the resulting signal-to-noise ratios (SNRs) and related integration times that are needed to perform imaging measurements under realistic conditions. It is known that deconvolution, removing the blur due to the SGL's spherical aberration substantially decreases the SNR. Our key finding is that this ``penalty'' is significantly mitigated when sampling locations in the image plane (image pixels) remain widely spaced. Consequently, we conclude that an imaging mission is challenging but feasible, using technologies that are either already available or in active development. Under realistic conditions, high-resolution imaging of Earth-like exoplanets in our galactic neighborhood requires only weeks or months of integration time, not years as previously thought: a high quality $1000\times 1000$ pixel image of an Earth-like planet at Proxima Centauri could be obtained with SNR$>$10 using approximately 14 months of integration time.
\end{abstract}

\begin{keywords}
gravitational lensing: strong $<$  Physical Data and Processes
\end{keywords}

\section{Introduction}
\label{sec:aintro}

The challenges of direct imaging of exoplanets are well known. Planets are small, very distant, and not self-luminous. They appear on top of a highly contaminated light background \citep{Traub:2010,Gaudi:2013}. Resolved imaging of such objects would require prohibitively large telescopes or interferometric baselines. For instance, consider an Earth-like exoplanet at a distance of $z_0=30$~pc. This object subtends an angle of $2R_\oplus/z_0=1.38\times 10^{-11}$~rad = $2.84\times 10^{-6}$~arcsec; to resolve it with just one pixel, one needs access to a diffraction-limited telescope with an aperture  of $d\sim 1.22 \lambda /(2R_\oplus/z_0)=88.53 ~(\lambda/1~\mu{\rm m})(z_0/30~{\rm pc})$~km, which is not practical. If $N$ linear pixels within the image are desired, then, again based on the diffraction limit, the aperture would have to scale as $N d$. Using optical interferometers for this purpose would not only involve variable and re-orientable interferometric baselines on the order of tens of kilometers but also telescope apertures of several tens of meters, each equipped with an external coronagraph \citep{Angel:2003} (e.g., starshade) to block out the light from the exoplanet's host star. Even with these parameters, interferometers would require integration times of hundreds of thousands to millions of years to reach a reasonable signal-to-noise ratio (SNR) of $\gtrsim 7$ to overcome the noise from exo-zodiacal light. As a result, direct resolved multipixel imaging of terrestrial exoplanets relying on conventional astronomical techniques and instruments is not feasible.

Motivated by these challenges, we considered the solar gravitational lens (SGL), as the means for direct imaging of exoplanets \citep{Turyshev:2017,Turyshev-Toth:2017}. The SGL results from the gravitational bending of light rays that propagate near the Sun. The SGL possesses truly impressive properties, including significant light amplification, $\mu_0=4\pi^2 r_g/\lambda \sim 1.17\times 10^{11} (1~\mu{\rm m}/\lambda)$, where $r_g$ is the solar Schwarzschild radius $r_g=2GM_\odot/c^2=2.95$ km; and an angular resolution of $\delta\theta=0.38\lambda/b\sim 1.03\times 10^{-10}(1~\mu{\rm m}/\lambda)(R_\odot/b)$ arcsec, where $b\gtrsim R_\odot$ is the impact parameter \citep{Turyshev-Toth:2020-extend}. Due to its impressive optical properties,  the SGL presents us with the only realistic means to overcome the challenges of resolved exoplanet imaging; it allows to do so by using technologies that are presently available or already in active development \citep{Turyshev-etal:2020-PhaseII,Helvajian-etal:2022}. The SGL offers a way to realize the  age-old human dream to see alien worlds that may exist on terrestrial exoplanets in our galactic neighborhood, especially worlds that could harbor life \citep{Seager:2010}.

To explore the imaging capabilities of the SGL, we extensively studied its optical properties \citep{Turyshev-Toth:2017,Turyshev-Toth:2019-extend,Turyshev-Toth:2019-blur,Turyshev-Toth:2019-image,Turyshev-Toth:2020-extend}. We conducted a series of numerical investigations of exoplanet imaging with the SGL \citep{Toth-Turyshev:2020}. That work allowed us to identify the basic properties of a solar coronagraph \citep{Shao-etal:2017,Zhou:2018} and model image deconvolution \citep{Shao:2018,Turyshev-etal:2018}. We were able to confirm the feasibility of using the SGL for imaging of faint sources such as distant exoplanets. We  considered the propagation of electromagnetic (EM) waves near the Sun and developed a Mie theory that accounts for the refractive properties of the gravitational field of the Sun and that of the free electron plasma in the extended solar system.
We have shown that the wavelength-dependent plasma effect is important at radio frequencies, where it drastically reduces both the amplification factor of the SGL and also its angular
resolution. However, for optical and shorter wavelengths, the plasma's contribution to the EM wave is negligibly small, leaving the plasma-free optical properties of the SGL practically unaffected \citep{Turyshev-Toth:2019,Turyshev-Toth:2019-extend,Turyshev-Toth:2019-corona}. These investigations focused only on the optical properties of the SGL and also on the light propagation through the plasma of the solar corona \citep{Turyshev-Toth:2019,Turyshev-Toth:2019-corona}, but not yet explicitly treating the solar corona brightness as a potential noise contribution.

Any use of the SGL requires observing the Einstein ring around the Sun, on the background of the solar corona. An initial analysis of the solar corona in the context of the SGL was offered by \cite{Landis:2016}. A more detailed investigation \citep{Willems:2018} identified the brightness of the solar corona as a significant source of noise that strongly affects imaging with the SGL by reducing the SNR and extending the per-pixel integration time. That study  suggested that the total time needed to recover megapixel images even from a nearby exoplanet is beyond the realm of practical mission durations.

Meanwhile, we developed a comprehensive formalism to investigate the imaging of extended sources with the SGL under realistic observing conditions \citep{Turyshev-Toth:2019,Turyshev-Toth:2019-blur,Turyshev-Toth:2019-image}. As a part of that effort, we developed an estimate for the SNR that includes an updated solar corona model, established the imaging geometry, and considered realistic observational scenarios \citep{Turyshev-Toth:2020-extend}. We validated these results using numerical simulations, including both direct deconvolution and deconvolution using the method of Fourier quotients. Consequently, we were able to explore imaging with the SGL, taking into account factors not considered by \citep{Willems:2018} such as pixel spacing in the image plane and various means to block the sun using either a coronagraph or a strashade \citep{Turyshev-Toth:2022-mono-SNR}. As a result, it became evident that under most observing scenarios, high-resolution imaging of exoplanets with the SGL remains manifestly feasible in the context of a realistic near-term space mission to the SGL focal region \citep{Turyshev-etal:2020-PhaseII,Toth-Turyshev:2020,Helvajian-etal:2022}.

Here we present a comprehensive summary of these efforts. Our objective is to evaluate and demonstrate the feasibility of using the SGL, for establishing realistic imaging scenarios and expectations for direct multipixel imaging of exoplanets within the context of a deep space mission that is based on available technology and executed in realistic timeframes. Previously published results included pessimistic estimates that questioned the feasibility of imaging with the SGL due to unrealistically long integration times even for nearby targets \citep{Willems:2018}, as well as overly optimistic results suggesting that a single snapshot of the Einstein ring taken from a single location on the image plane might be sufficient to reconstruct intermediate resolution images of even very distant targets \citep{MM:2022}. Reality is somewhere in between: the penalty incurred by image deconvolution is not as severe when one accounts for realistic pixel spacing, but the stochastic noise due to the solar corona is significant and must not be ignored or underestimated in any practical imaging scenario.

Our paper is organized as follows:
Section~\ref{sec:opt-prop} introduces the SGL and discusses its optical properties relevant to imaging of exoplanets.
In Section \ref{sec:sol-corona} we discuss the solar coronagraph that the relevant signal from the solar corona.
In Section~\ref{sec:snr-c} we calculate the signal to noise ratio and the effects of deconvolution, and compare results to numerical simulations.
In Section \ref{sec:disc} we discuss the results and avenues for the next phase of our investigation of the SGL.

\section{Summary of the optical properties of the SGL}
\label{sec:opt-prop}

We consider an exoplanet as an extended source of radius $R_\oplus$, located at a large, but finite distance $z_0$ from the Sun, using an imaging geometry summarized in Fig.~2 of \citep{Turyshev-Toth:2020-extend}.
The image of this object is formed in the strong interference region of the SGL, at the heliocentric position of $\overline z\geq b^2/2r_g=547.76\, (b/R_\odot)^2$ astronomical units (AU). There is no single focal point of the SGL, but a semi-infinite focal line. The SGL acts as a convex lens with negative spherical aberration. It compresses the source, forming the image of an exoplanet within the volume occupied by a cylinder with a diameter of $2r_\oplus=(\overline z/z_0) 2R_\oplus \simeq1.34 \,(\overline z/650\,{\rm AU})(30\,{\rm pc}/z_0)\,{\rm km}$. Placing a spacecraft in any of the image planes\footnote{Note that in the field of strong gravitational lensing, the ``image plane'' of the lensing system is defined as the primary lens plane \citep{Schneider-Ehlers-Falco:1992}, whereas the ``image plane'' used in this manuscript is considered to be the observer's location in the SGL's strong interference region in the vicinity of the primary optical axis, as discussed in \citep{Turyshev-Toth:2020-extend}.
}  within this cylinder allows to take data that may be used to assemble an image of the distant, faint target \citep{Toth-Turyshev:2020}.

We consider resolved imaging of extended sources with the SGL with a modest-size telescope with aperture $d\ll 2r_\oplus$ (see details in \citep{Toth-Turyshev:2020,Turyshev-Toth:2020-extend}). The light collected by the telescope is the sum of light from the ``directly imaged'' region of the source and light from the rest of the planet. The directly imaged region on the source is the area with the diameter of $D=(z_0/ \overline z)d=9.5\,(z_0/30\,{\rm pc})(650\,{\rm AU}/\overline z)(d/1\,{\rm m})\,{\rm km}$. This directly imaged region is $2R_\oplus/D=1340$ times smaller than the rest of the planet.

This difference between the sizes of the directly imaged region and the entire planet is important for signal estimation.  As was shown \citep{Turyshev-Toth:2017}, the point-spread function (PSF) of the SGL has the form $\propto J^2_0(k \rho\sqrt{2r_g/\overline z})$, where $\rho$ is the deviation from the optical axis. For large $\rho$, this PSF behaves on average as $\propto 1/\rho$, which is different from the typical PSF of a thin lens that is given by $\propto(2J_1(\alpha \rho)/(\alpha \rho))^2$, which, for large $\rho$, behaves on average as $\propto 1/\rho^3$. The $1/\rho$ behavior of the SGL PSF implies that the telescope, although it points toward the directly imaged region, collects more light from areas far from this region. This extra signal from the rest of the exoplanet results in the emergence of blur with an intensity that may overwhelm any light received from the directly imaged region.

As the SGL PSF is known, the signal from the directly imaged region can be recovered in principle. The mapping between the source and the blurred image is linear and invertible. Deconvolution, in principle, amounts to subtracting the contribution of the blur, leaving us with only the signal of interest at each image pixel. Deconvolution can also be performed in Fourier space, leading to a significant improvement in computational efficiency, essentially making it possible to carry out deconvolution even of megapixel-resolution images without specialized computational resources, on ordinary desktop computers.

The downside of deconvolution is that it increases noise \citep{Toth-Turyshev:2020}. This can be understood easily in principle when we consider that deconvolution amounts to removing, from the observed signal, the contributions due to blur. However, stochastic (Gaussian or Poisson) noise by its nature is always additive. Therefore, the deconvolution process reduces the signal even as it increases noise.

Clearly, sampling frequencies -- both spatial and temporal -- affect the quality of the recovered image.  Ultimately, image quality depends on the SNR achieved per image pixel for each position of the imaging telescope on the image plane. Our goal is to present a quantitative analysis of the resulting SNR, backed by results from computer simulations.

To describe the image of faint objects with the SGL, we take an imaging telescope and position it in the image plane at the strong interference region of the SGL. Looking back at the Sun, this imaging telescope sees an Einstein ring containing light from both the directly imaged region (contributing equally to the entire Einstein ring) and other regions of the exoplanet (contributing light to various segments of the Einstein ring). Most of the Einstein ring seen by the telescope is from light that comes from this blur, contributions from regions of the exoplanet other than the directly imaged region.

\subsection{Signal from an exoplanet}
\label{sec:sig-exop}

A telescope with a modest aperture, $d\ll r_\oplus$,  traversing the image plane, will receive signals from different parts of the exoplanet contributing various quantities of blur.  As was shown by \cite{Turyshev-Toth:2019-blur,Turyshev-Toth:2020-extend}, the power of the received signal at the focal plane of the imaging telescope is a function of the telescope's position on the image plane. For a uniform source brightness, $B_{\tt s}$, the result depends only on the separation from the optical axis, $\rho_0$. The power received from the directly imaged region on a resolved exoplanet for $\rho_0=0$  and measured along the image of the Einstein ring that forms in the focal plane of a diffraction-limited telescope  is given as
{}
\begin{eqnarray}
P_{\tt fp.dir}&=& \epsilon_{\tt dir} B_{\tt s} \frac{\pi^2 d^3}{4{\overline z}}\sqrt{\frac{2r_g}{\overline z}},
  \label{eq:pow**}
\end{eqnarray}
where  $\epsilon_{\tt dir} =0.77$ is the encircled energy fraction for directly imaged region, see \cite{Turyshev-Toth:2020-extend}.  The power (\ref{eq:pow**}) is independent of the observing wavelength and the distance to the target; however it is a strong function of the telescope's aperture.

At the same time, the power at the Einstein ring at the detector placed in the focal plane of an optical telescope is dominated by the blur and is given as
{}
\begin{eqnarray}
P_{\tt fp.blur}(\rho_0)&=&
 \epsilon_{\tt blur} B_{\tt s}\pi^2 d^2  \frac{R_\oplus}{2z_0}\sqrt{\frac{2r_g}{\overline z}}\,\mu(\rho_0),
\qquad{\rm with}\qquad
\mu(r_0)=
\begin{cases}
\epsilon(\rho_0),&0\leq \rho_0/r_\oplus\leq 1,\\
\beta(\rho_0),&\rho_0/r_\oplus\geq 1,
\end{cases}
  \label{eq:Pexo-blur}
\end{eqnarray}
where $\epsilon_{\tt blur}=0.69$ is the encircled energy fraction for the rest of the planet \citep{Turyshev-Toth:2020-extend} and the factors $\epsilon(\rho_0)$ and $\beta(\rho_0)$ are
{}
\begin{eqnarray}
\epsilon(\rho_0)
~=~\frac{1}{2\pi}\int_0^{2\pi} \hskip -3pt d\phi\sqrt{1-\Big(\frac{\rho_0}{r_\oplus}\Big)^2\sin^2\phi}=\frac{2}{\pi}{\tt E}\Big[\frac{\rho_0}{r_\oplus}
\Big],
\label{eq:eps_r0}
\qquad
\beta (\rho_0)
~=~\frac{1}{\pi}\int_{\phi_-}^{\phi_+} \hskip -3pt d\phi\sqrt{1-\Big(\frac{\rho_0}{r_\oplus}\Big)^2\sin^2\phi}=\frac{2}{\pi}{\tt E}\Big[\arcsin \frac{r_\oplus}{\rho_0},
\frac{\rho_0}{r_\oplus}
\Big],
\label{eq:beta_r0}
\end{eqnarray}
where ${\tt E}[x]$ is the elliptic integral and  ${\tt E}[a,x]$ is the incomplete elliptic integral \citep{Abramovitz-Stegun:1965}. The behavior of $\epsilon(\rho_0)$ and $\beta(\rho_0)$  is shown in Fig.~6 of \cite{Turyshev-Toth:2020-extend}. The blur's contribution in (\ref{eq:Pexo-total}) is captured by the factor $\mu(\rho_0)$, which, outside the directly projected image of the exoplanet, falls off as $\propto 1/\rho_0$, as expected from the PSF of the SGL.

As a result, the total power received by the telescope is $P_{\tt exo}(\rho_0)=P_{\tt fp.dir}(\rho_0)+P_{\tt fp.blur}(\rho_0) \simeq P_{\tt fp.blur}(\rho_0)$. In other words, the total power is given by the following expression:
{}
\begin{eqnarray}
P_{\tt exo}(\rho_0)= \epsilon_{\tt blur} B_{\tt s}\pi^2 d^2  \frac{R_\oplus}{2z_0}\sqrt{\frac{2r_g}{\overline z}}\,\mu(\rho_0).
  \label{eq:Pexo-total}
\end{eqnarray}
Next,  we express the surface brightness via familiar quantities:
{}
\begin{eqnarray}
B_s=\frac{g}{\pi}\alpha I_0,  ~~~[{\rm W}/{\rm m}^{2} {\rm sr}],
  \label{eq:Bs}
\end{eqnarray}
where $\alpha$ is the source's albedo, $I_0$ is the incident radiant energy emitted by the host star that is received at the top of the exoplanetary atmosphere (instellation) in units of ${\rm W}/{\rm m}^2$. The factor $g$ determines the surface properties of the reflector: $g=\frac{2}{3}$ for a Lambertian surface.
Overall, the brightness $B_s$ has the dimensions of $[{\rm W}/{\rm m}^{2} {\rm sr}]$.

\begin{table*}
\caption{Signal received from an exoplanet.
\label{tb:signal}}
\begin{tabular}{|l|c|c|c|c|}\hline
Parameter  &Symbol & from \cite{Willems:2018}&  \multicolumn{2}{c|}{from \cite{Turyshev-Toth:2020-extend}}\\\hline\hline
Encircled energy & $\epsilon_{\tt blur} $ & 1 &\multicolumn{2}{c|}{0.69}\\
Reflection factor & $g$ & $\frac{1}{2}$ & \multicolumn{2}{c|}{$\frac{2}{3}$}\\
Planetary albedo & $\alpha$ &0.3 & \multicolumn{2}{c|}{0.3}\\
Solar irradiance, [W/m$^2$] & $I_0$&1,000 & \multicolumn{2}{c|}{1,366.83}\\
Telescope diameter, [m]& $d$& 1 & \multicolumn{2}{c|}{1}\\
Explanent's radius,   [km]   & $R_\oplus$ & 6,370 &\multicolumn{2}{c|}{6,371}\\
Distance to exoplanet, [pc]& $z_0$ &1.3 &1.3 &30 \\
Telescope's position, [AU]&${\overline z}$& 1,200 &1,200 &650 \\
Blur envelope factor & $\mu(\rho_0)$~\, &1 & \multicolumn{2}{c|}{$\mu(\rho_0)$}\\\hline\hline
Power received, [W] &$P_{\tt exo}(\rho_0)$ & $2.14\times 10^{-13}$ &  $2.70\times 10^{-13}$ & $1.59\times 10^{-14}$\\
Photon flux received, [phot/s]& $Q_{\tt exo}(\rho_0)$ & $1.08\times 10^{6}$  & $1.36\times 10^{6}$  & $8.01\times 10^{4}$ \\
\hline
\end{tabular}
\end{table*}

Using (\ref{eq:Bs}) in (\ref{eq:Pexo-total}), we have the following expressions for the power, $P_{\tt exo}(\rho_0)$, and photon flux, $Q_{\tt exo}(\rho_0)$, received by the telescope
{}
\begin{eqnarray}
P_{\tt exo}(\rho_0)&=&  \epsilon_{\tt blur} g\alpha I_0 {\pi d^2}\frac{R_\oplus}{2z_0}\sqrt{\frac{2r_g}{\overline z}}\mu(\rho_0)= \nonumber\\
&=&1.59\times 10^{-14}\mu(\rho_0)\Big(\frac{ \epsilon_{\tt blur}}{0.69}\Big)\Big(\frac{g}{2/3}\Big)\Big(\frac{I_0}{1366.83}\Big)
\Big(\frac{d}{1\,{\rm m}}\Big)^2\Big(\frac{650\,{\rm AU}}{\overline z}\Big)^\frac{1}{2}\Big(\frac{30\,{\rm pc}}{z_0}\Big)~{\rm W},
\label{eq:blur-sP} \\
Q_{\tt exo}(\rho_0)&=&\frac{\lambda}{hc}P_{\tt exo}(\rho_0)= \nonumber\\
&=&8.01\times 10^4\mu(\rho_0)
\Big(\frac{ \epsilon_{\tt blur}}{0.69}\Big)\Big(\frac{g}{2/3}\Big)\Big(\frac{I_0}{1366.83}\Big)\Big(\frac{d}{1\,{\rm m}}\Big)^2\Big(\frac{650\,{\rm AU}}{\overline z}\Big)^\frac{1}{2}\Big(\frac{30\,{\rm pc}}{z_0}\Big)
\Big(\frac{\lambda}{1\,\mu{\rm m}}\Big)~{\rm phot/s}.~~~~~
\label{eq:blur-s}
\end{eqnarray}

Although derived using a different approach (see details in \cite{Turyshev-Toth:2020-extend}),
the analytical expression for result (\ref{eq:blur-sP}),  is formally identical to that published in \cite{Willems:2018}. Table~\ref{tb:signal} compares numerical estimates for an exoplanet located at 1.3~pc and observed through the SGL using a 1-m telescope at 1200~AU. The estimated signal levels are comparable in magnitude, with any differences due to the estimated values of $\epsilon_{\tt blur}$, $g$ and $I_0$. For comparison, a third case (exoplanet at 30~pc, telescope at 650~AU) is also shown.\footnote{Using (\ref{eq:blur-s}), we can see that in the absence of noise, which is the case considered by \cite{MM:2022}, the attainable signal-to-noise ratio is ${\rm SNR}\simeq \sqrt{Q_{\tt exo}(\rho_0)}\sim 280$ in 1 second of integration time for a target at 30~pc, or $\sim155$ when the target is at $z_0=100$~pc. Achieving the desired SNR of $4\times 10^3$ or $2\times 10^5$, mentioned by \cite{MM:2022}, would require $\sim$11 minutes and $\sim$5.3 years, respectively, again, in the complete absence of corona or other sources of noise, considering only the stochastic nature of the signal that consists of discrete photons.}

\section{Solar coronagraph and signal from the solar corona}
\label{sec:sol-corona}

\subsection{Solar coronagraph}
\label{sec:coronagr}

Solar coronagraphy, invented by \cite{Lyot:1932}, is used to study the solar corona and other celestial objects that appear in the vicinity of the Sun.  The goal is to reproduce solar eclipses artificially, blocking light from the Sun. Solar coronagraphs differ from the coronagraphs that are used in attempts to resolve exoplanets in distant solar systems. In those cases, the objective is to block out light from the host star, a point source. In contrast, a coronagraph for the  SGL must work in the tradition of Lyot's original coronagraph, as it must block light from the Sun, while allowing light in from the Einstein ring that appears on the background of the solar corona, barely separated from the solar disk.

\begin{figure}
\centering\includegraphics[scale=0.833]{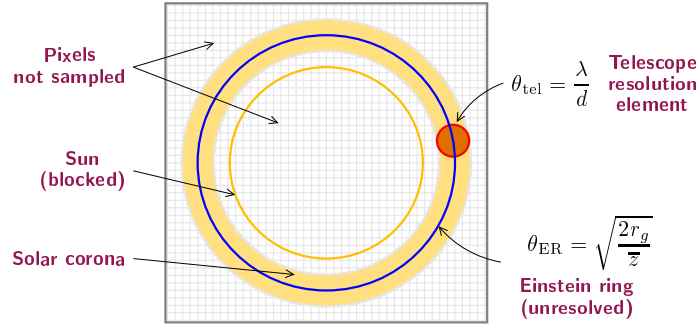}\par
\caption{\label{fig:foc-plane-det}
The Einstein ring in the imaging telescope's sensor plane.
Light is not collected from pixels either within and outside the Einstein ring. The thickness of the exposed area is determined by the diffraction limit of the optical telescope at its typical observational wavelength.}
\end{figure}

The already available design for the SGL coronagraph \citep{Zhou:2018} rejects solar light with a contrast ratio of $\sim 10^{-7}$, which is sufficient for our purposes. At this level of rejection, the light from the solar disk is reduced to the level comparable to the brightness of the solar corona. Additional unwanted light from the corona itself can be rejected by implementing an annular mask in the telescope's focal (sensor) plane. This amounts to collecting light only from the Einstein ring of the intended target and ignoring sensor pixels inside or outside the ring. This approach directly translates into a reduction of stochastic noise and a consequent decrease in the required integration time to achieve a desired SNR.

\subsection{Signal from the solar corona}
\label{sec:sol-corona-model}

As we established, the Einstein ring that forms around the Sun from light emitted by the exoplanet that is the observational target appears on the bright background of the solar corona. Even if we assume that the corona background can be accurately estimated (or measured by other instruments) and removed, as light is quantized into photons, inevitably, there is stochastic noise in the form of Poisson (approximately Gaussian) shot noise.

To assess this noise contribution by the corona, we need to estimate the signal from the solar corona within the annulus that corresponds to the Einstein ring around the Sun, formed by light from the target exoplanet. In the region occupied by the image of the Einstein ring in the focal plane of a diffraction-limited telescope, the corona contribution is given after the coronagraph as\footnote{The upper integration limit for the coronagraph was reduced from $\infty$ to $2\theta_0$ in \cite{Willems:2018}, but this does not appear to have physical justification, so we continue to use the more conservative results obtained from (\ref{eq:pow-fp=+*}), yielding a corona contribution approximately 14\% more than estimated in \cite{Willems:2018}.}:
{}
\begin{eqnarray}
P_{\tt
cor}
&=&\epsilon_{\tt cor}\,\pi({\textstyle\frac{1}{2}}d)^2
\int_0^{2\pi}\hskip -4pt d\phi
\int_{\theta_0}
^\infty\hskip -4pt \theta d\theta \, B_{\tt cor}(\theta),
  \label{eq:pow-fp=+*}
\end{eqnarray}
where $\theta=\rho/{\overline z}$ and $\theta_0=R_\odot/{\overline z}$.

The surface brightness of the solar corona $B_{\tt cor}(\theta)$ is taken from  the Baumbach model \citep{Baumbach:1937,Golub-Pasachoff-book:2017}:
{}
\begin{align}
B_{\tt cor}(\theta)&= 18.9\Big[2.565 \Big(\frac{\theta_0}{\theta}\Big)^{17}+1.425\Big(\frac{\theta_0}{\theta}\Big)^{7}+ 5.32\times 10^{-2} \Big(\frac{\theta_0}{\theta}\Big)^{2.5}\Big]  ~~   \frac{\rm W}{{\rm m}^2\,{\rm sr}}.
\label{eq:model-th-w}
\end{align}

Another, more recent and a bit more conservative model \cite{November:1996} is given by:
{}
\begin{align}
B_{\tt cor}(\theta)&{}= 20.09\Big[3.670 \Big(\frac{\theta_0}{\theta}\Big)^{18}+1.939\Big(\frac{\theta_0}{\theta}\Big)^{7.8}+ 5.51\times 10^{-2} \Big(\frac{\theta_0}{\theta}\Big)^{2.5}\Big]  ~~   \frac{\rm W}{{\rm m}^2\,{\rm sr}}.\label{eq:model-th0}
\end{align}

Using the Baumbach model and the value of $\epsilon_{\tt cor} = f = 0.35$ , as it was done by \cite{Willems:2018}, yields the following estimate for the photometric signal received from the solar corona:
 {}
\begin{align}
P^{\tt }_{\tt cor}&{}=
2.76\times 10^{-10}\Big(\frac{d}{1\,{\rm m}}\Big)^2\Big(\frac{\rm 1200 \,AU}{\overline z}\Big)^2~{\rm W},
  \label{eq:phot3}\\
Q^{\tt }
_{\tt cor}&{}=
\frac{\lambda}{hc}P^{\tt }_{\tt cor}=
1.39\times 10^9
\Big(\frac{d}{1\,{\rm m}}\Big)^2\Big(\frac{\rm 1200 \,AU}{\overline z}\Big)^2\Big(\frac{\lambda}{1\,\mu{\rm m}}\Big)~{\rm photons/s}.
\label{eq:blur-s-wf}
\end{align}

Using (\ref{eq:model-th0}) for the nominal mission range of $\overline z=650$~AU, as in \cite{Turyshev-Toth:2020-extend}, we obtain
{}
\begin{align}
P_{\tt cor}=1.23\times 10^{-9}
\Big(\frac{d}{1\,{\rm m}}\Big)^2\Big(\frac{\rm 650 \,AU}{\overline z}\Big)^2~{\rm W},
\end{align}
where we use $\epsilon_{\tt cor}=0.36$, computed in \cite{Turyshev-Toth:2020-extend} as the fraction of the encircled energy for the solar corona and also consistent within rounding with the value $f=0.35$ in \cite{Willems:2018}. This corresponds to the corona photon flux of
{}
\begin{align}
Q_{\tt cor}=
\frac{\lambda}{hc}P^{\tt }_{\tt cor}=
6.20\times 10^9\Big(\frac{d}{1\,{\rm m}}\Big)^2\Big(\frac{\rm 650 \,AU}{\overline z}\Big)^2\Big(\frac{\lambda}{1\,\mu{\rm m}}\Big)~{\rm photons/s}.
\label{eq:pow-fp=+*4+2}
\end{align}

For the same heliocentric ranges, the two models are within $\sim 30$\% of each other. These estimates will need to be refined by analyzing the spectral composition of the corona and signal as was recently done in \citep{Turyshev-Toth:2022-mono-SNR}.

\section{Signal to noise ratio in the presence of the corona}
\label{sec:snr-c}

\subsection{Signal to noise ratio of the observed image}

Assuming that the contribution of the solar corona is removable (e.g., by observing the corona from a slightly different vantage point) and only stochastic (shot) noise remains, we estimate the resulting signal-to-noise ratio as
\begin{align}
{\rm SNR}_{\tt C}=\frac{Q_{\tt exo}}{\sqrt{Q_{\tt exo}+Q_{\tt cor}}},
\end{align}
in the regime dominated by the solar corona. The subscript ${\tt C}$ signifies that it is a signal that has been convolved by the SGL's PSF.

Using values consistent with \cite{Willems:2018} (see Table~\ref{tb:signal}) in the expressions for the signal from the exoplanet  at $z_0=1.3$~pc (\ref{eq:blur-s}) (where it was taken only at the center of the image, $\rho_0=0$, thus, $\mu(\rho_0)=1$) and the solar corona (\ref{eq:blur-s-wf}), for ${\overline z}=1200$~AU we obtain
the following ${\rm SNR}_{\tt C}$ of the convolved image:
{}
\begin{equation}
{\rm SNR}_{\tt C}=
36.44\,\Big(\frac{d}{1\,{\rm m}}\Big)\Big(\frac{\overline z}{1200\,{\rm AU}}\Big)^\frac{1}{2}\Big(\frac{1.3\,{\rm pc}}{z_0}\Big)\Big(\frac{\lambda}{1\,\mu{\rm m}}\Big)^\frac{1}{2}\,\sqrt{\frac{t}{1\,{\rm s}}}.
\label{eq:snr-cor-W}
\end{equation}

In \cite{Turyshev-Toth:2020-extend}, this SNR was estimated by assuming slightly different values in (\ref{eq:blur-s}), see Table~{\ref{tb:signal}. We estimated the ${\rm SNR}$ of the convolved image for an exoplanet at $z_0=30$~pc and the telescope at heliocentric distance of ${\overline z}=650$~AU. This quantity is easy to compute from (\ref{eq:blur-s}) and (\ref{eq:pow-fp=+*4+2}) in the following form:
{}
\begin{align}
{\rm SNR}_{\tt C}(\rho_0)&{}=1.02\mu(\rho_0)
\Big(\frac{d}{1\,{\rm m}}\Big)\Big(\frac{\overline z}{650\,{\rm AU}}\Big)^\frac{1}{2}\Big(\frac{30\,{\rm pc}}{z_0}\Big)\Big(\frac{\lambda}{1\,\mu{\rm m}}\Big)^\frac{1}{2}\,\sqrt{\frac{t}{1\,{\rm s}}}.
\label{eq:snr-cor}
\end{align}

Using (\ref{eq:snr-cor}), we  may evaluate the SNR under the same conditions, for $\rho_0=0$, $z_0=1.3$~pc and ${\overline z}=1200$~AU and obtain ${\rm SNR}_{\tt C}=31.9$, which is consistent with (\ref{eq:snr-cor-W}).

\subsection{SNR of the deconvolved image}

To estimate the impact of deconvolution on the signal-to-noise ratio, consistent with \cite{Willems:2018} and \cite{Turyshev-Toth:2020-extend} we model the exoplanet as a source with uniform brightness and represent the image as a collection of $N$ pixels. We denote the post-deconvolution signal-to-noise ratio as ${\rm SNR}_{\tt R}$.

In \cite{Turyshev-Toth:2020-extend} we estimated the deconvolution penalty as
\begin{align}
\frac{{\rm SNR}_{\tt R}}{{\rm SNR}_{\tt C}}=
\frac{\displaystyle\frac{1}{N}\sum_{i=1}^N\sum_{j=1}^N\, C^{-1}_{ij}}{\Big({\displaystyle\frac{1}{N}\sum_{i=1}^N\sum_{j=1}^N\,(C^{-1}_{ij})^2}\Big)^\frac{1}{2}~},
\label{eq:penalty0}
\end{align}
where $C_{ij}$ is the convolution matrix and $N$ is the total number of image pixels. To estimate the inverse of the convolution matrix, we approximated it in the form
\begin{align}
C_{ij}\sim\tilde{C}_{ij}=\frac{4}{\pi\alpha d}\Big(\mu\delta_{ij}+\nu U_{ij}\Big),\label{eq:Cij}
\end{align}
where $\delta_{ij}$ is the Kronecker-delta, $U_{ij}$ is the everywhere-one matrix, $\nu\ll 1$ and $\mu=1-\nu$.

We can estimate $\nu$ using the averaged SGL PSF and replacing summations over large $N$ with integrals:
\begin{align}
\nu=\frac{1}{A}\iint\limits_{x,y=-\sqrt{N}d/2}^{\sqrt{N}d/2} dxdy\frac{d}{4\sqrt{x^2+y^2}}=\frac{4}{\pi}\frac{\ln(\sqrt{2}+1)}{\sqrt{N}},
\label{eq:nu}
\end{align}
where the light collecting area is given by $A\sim N\pi\frac{1}{4}d^2$, consistent with our PSF estimates assuming a telescope with a circular aperture used to traverse the image plane and sample image pixels.

Using this approximation allowed us to explicitly calculate $\tilde{C}_{ij}^{-1}$ and thus estimate the deconvolution penalty for large $N$, $\mu\sim 1$, as
{}
\begin{eqnarray}
\frac{{\rm SNR}_{\tt R}}{{\rm SNR}_{\tt C}}=
\frac{\displaystyle\frac{1}{N}\sum_{i=1}^N\sum_{j=1}^N\, \widetilde{C}^{-1}_{ij}}{\Big({\displaystyle\frac{1}{N}\sum_{i=1}^N\sum_{j=1}^N\,(\widetilde{C}^{-1}_{ij})^2}\Big)^\frac{1}{2}~}
= \frac{\mu}{\nu N}\sim\frac{0.891}{\sqrt N}.
\label{eq:penalty}
\end{eqnarray}
This result describes image deconvolution using adjacent pixels.

When pixels are not adjacent, however, $\nu$ is scaled by $d/D$ where we denote the pixel spacing by $D$. Thus we obtain the following estimate for the deconvolution penalty:
{}
\begin{align}
\frac{{\rm SNR}_{\tt R}}{{\rm SNR}_{\tt C}}\sim 0.891\frac{D}{d\sqrt{N}}.
\label{eq:penalty2}
\end{align}

This result can yield significant improvement in the SNR and a corresponding reduction in integration times when imaging either exoplanets in nearby systems or more distant exoplanets at reduced resolution.

\begin{figure}
\centering\includegraphics{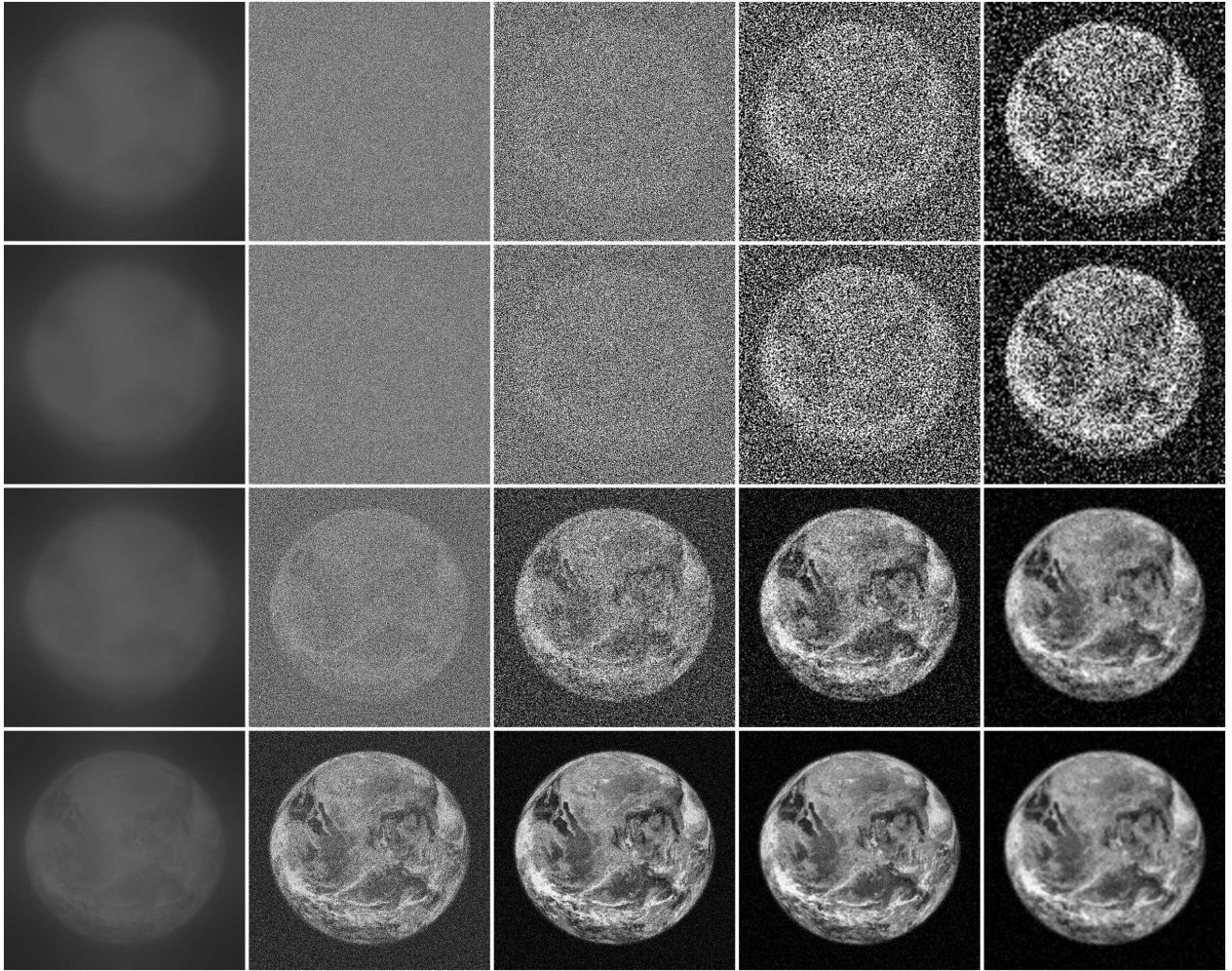}\par
\caption{\label{fig:earth-m}Deconvolution results using the Earth as a stand-in for an exoplanet at a variety of distances and resolutions as characterized by the projected size of the exoplanet image in the SGL image plane: first row, 1024~m (resulting in adjacent pixels at the maximum resolution); second row, 1300~m (corresponding to an exo-Earth at 30~pc with the observing telescope at 650~AU); third row, 10000~m (exo-Earth at 4~pc); and last row, 57,000~m (exo-Earth at 1.3~pc with the observing telescope at 1200~AU from the Sun). Deconvolution was performed after Gaussian noise was added to the convolved image (first column) at ${\tt SNR}_{\rm C}=10$. Columns 2--5 show results at $1024\times 1024$, $512\times 512$, $256\times 256$ and $128\times 128$ pixel resolutions.}
\end{figure}

\begin{table*}
\caption{\label{tb:earth-m}Comparison of the estimated deconvolution penalty to the penalty obtained from simulation, using four different image sizes. A 57000~m diameter image corresponds to an exo-Earth at 1.3 pc observed with an SGL instrument at 1200~AU; 10000~m corresponds to an exo-Earth at 4~pc, observed from 650~AU; 1300~m corresponds to the same exo-Earth at 30~pc; finally, the 1024~m image is included for comparison, as a reference case with adjacent pixels at the maximum resolution.}
\begin{tabular}{c|cc|cc|cc|cc}
\multicolumn{1}{r}{Size:}&\multicolumn{2}{c|}{57000~m}&\multicolumn{2}{c|}{10000~m}&\multicolumn{2}{c|}{1300~m}&\multicolumn{2}{c}{1024~m}\\\hline
Resolution&sim.&est.&sim.&est.&sim.&est.&sim.&est.\\\hline\hline
$\phantom{0}128\times 128\phantom{0}$
 & 0.799 & $>1$
 & 0.411 & 0.544
 & $8.26\times 10^{-2}$ & $7.07\times 10^{-2}$
 & $8.25\times 10^{-3}$ & $6.96\times 10^{-3}$
\\
$\phantom{0}256\times 256\phantom{0}$
 & 0.498 & 0.775
 & 0.148 & 0.140
 & $2.16\times 10^{-2}$ & $1.77\times 10^{-2}$
 & $4.24\times 10^{-3}$ & $3.48\times 10^{-3}$
\\
$\phantom{0}512\times 512\phantom{0}$
 & 0.198 & 0.194
 & 0.042 & 0.034
 & $5.27\times 10^{-3}$ & $4.42\times 10^{-3}$
 & $2.03\times 10^{-3}$ & $1.74\times 10^{-3}$
\\
$          1024\times1024           $
 & 0.0581 & 0.0484
 & 0.0106 & 0.0085
 & $1.18\times 10^{-3}$ & $1.10\times 10^{-3}$
 & $8.56\times 10^{-4}$ & $8.70\times 10^{-4}$
\\\hline
\end{tabular}
\end{table*}

Numerical simulations using a variety of images, resolutions, and pre-deconvolution Gaussian noise consistently confirm that the penalty is indeed proportional to $1/\sqrt{N}$ with a numerical coefficient of $\sim 0.891D/d$ across a wide range of resolutions, image sizes and image patterns. A few representative cases are shown in Fig.~\ref{fig:earth-m}, using an image of the Earth as a stand-in. The corresponding deconvolution penalties are summarized in Table~\ref{tb:earth-m}.

As a general observation, we find that the simulation yields deconvolution SNRs that are similar to, or slightly better than the value predicted by Eq. (\ref{eq:penalty2}). This is, of course, not unexpected: the grossly simplified form of the convolution matrix, (\ref{eq:Cij}), is intended to capture the qualitative features of $C_{ij}$ in invertible form, but it is not accurate. However, the fact that the simulated results deviate only slightly from Eq.~(\ref{eq:penalty2}) and when they do, the estimate proves conservative, gives us strong confidence that (\ref{eq:penalty2}) reliably captures the SNR cost of deconvolution. The prediction fails only when the value it predicts approaches of exceeds 1. This, of course, is self-evident: deconvolution will not remove noise, it only redistributes it.

A perhaps striking aspect of these results is that the deconvolution penalty does not depend on the image content. This, however, is best understood if we realize that convolution is ultimately a linear process. Given a signal $s_i$ of $N$ pixels to which noise $\sigma_i$ is added, $s'_i=s_i+\sigma_i$, deconvolution yields $C^i_js'_i=C^i_j(s_i+\sigma_i)=C^i_js_i+C^i_j\sigma_i$, i.e., the impact of deconvolution on the noise does not depend on the signal.

For further details on the range of images, resolutions, and Gaussian noise that were used to validate our results, and a corresponding range of representative images see Appendix~\ref{appA}.

\subsection{Estimating the required integration time}

\begin{figure}
\centering\includegraphics{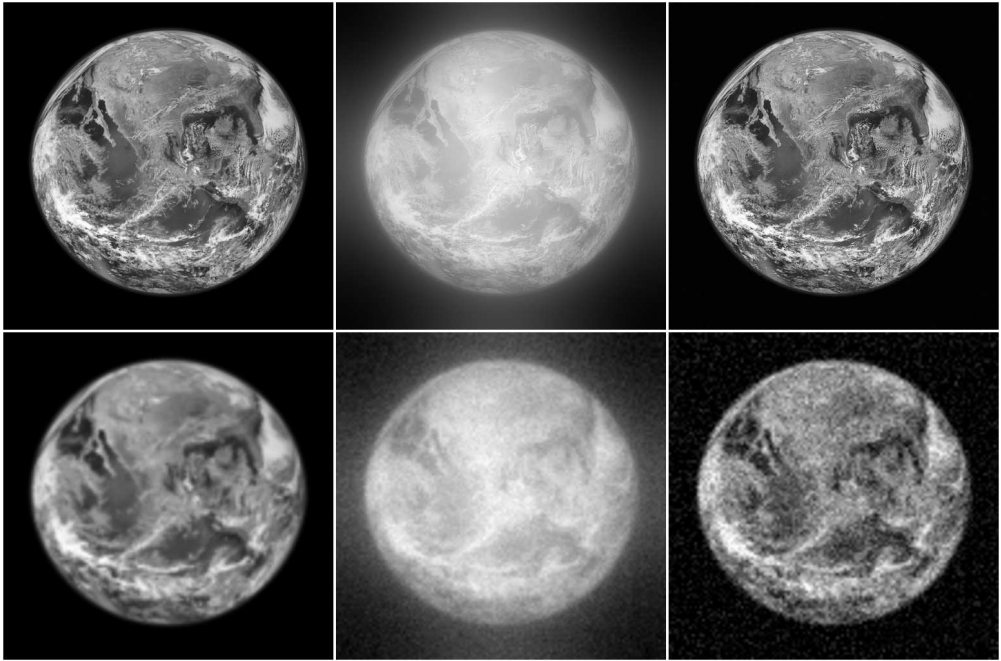}\par
\caption{\label{fig:earth1024}Top row: Simulated monochromatic imaging of an exo-Earth at $z_0=1.3$~pc from ${\overline z}=1200$~AU at $N=1024\times 1024$ pixel resolution using the SGL. Left: the original image. Middle: the image convolved with the SGL PSF, with noise added at ${\rm SNR}_{\rm C}=187$, consistent with a total integration time of $\sim$1.2 years. Right: the result of deconvolution, yielding an image with ${\rm SNR}_{\rm R}=11.4$. Bottom row: As above, but the exo-Earth is now at $z_0=30$~pc, imaged at $N=128\times 128$ pixel resolution, with noise added at ${\rm SNR}_{\rm C}=41.3$, consistent with an integration time of $\sim$0.85 years. The post-deconvolution SNR is ${\rm SNR}_{\rm R}=3.5$.}
\end{figure}

Given a target ${\rm SNR}_{\tt R}$ after deconvolution, we can estimate the pre-deconvolution value of ${\rm SNR}_{\tt C}$ and consequently, the required integration time by solving for $t$ in Eqs.~(\ref{eq:snr-cor-W}) or (\ref{eq:snr-cor}). For a target ${\rm SNR}_{\rm R}=10$, using $D=d$, $N=10^6$, from (\ref{eq:penalty2}), we have ${\rm SNR}_{\rm C}=1.12 \times 10^4$. Then, for an exoplanet at 1.3~pc and with the telescope at 1200~AU, from (\ref{eq:snr-cor-W}), we obtain the total integration time of $Nt=9.46\times 10^{11}~{\rm s}=3,000$~years, which is prohibitively long. However, it is important to note that the image of an exo-Earth at 1.3~pc, projected to an image plane at 1200~AU from the Sun, is almost 60~km wide. Therefore, in case a megapixel image is sought, individual image plane pixels will not be adjacent: they will be $D\sim 60$~meters apart. Consequently, the integration time penalty for a $d=1$ will scale dramatically, by a factor of $(d/D)^2=1/3600$, to approximately one year even when the more conservative corona model (\ref{eq:pow-fp=+*4+2}) is used.

To obtain the best estimate consistent with our nominal mission baseline with a telescope at 650~AU and an exo-Earth at 30 pc from the Earth, we use (\ref{eq:snr-cor}) and set $\mu(\rho_0)\sim 1$, characteristic of the image interior. The total integration time  needed to obtain an image of $N$ pixels is
\begin{align}
T&{}= 1.21\, {N^2{\rm SNR}_{\tt R}^2}
\Big(\frac{1~{\rm m}}{D}\Big)^2
\Big(\frac{z_0}{30~{\rm pc}}\Big)^2
\Big(\frac{650~{\rm AU}}{{\overline z}}\Big)
\Big(\frac{1~\mu{\rm m}}{\lambda}\Big)~~~{\rm s}.
\end{align}

Using $D=60$~m, $z_0=1.3$~pc, ${\overline z}=1200$~AU, for ${\rm SNR}_{\tt R}=10$ and $N=10^6$ we obtain $T\sim 1.2$~years. For comparison, at the nominal mission limit of $z_0=30$~pc, ${\overline z}=650$~AU, a good quality image of an exo-Earth with ${\rm SNR}_{\tt R}=3$, using $N=128\times 128$ image pixels that yield the pixel spacing of $D=2r_\oplus/128=10.47$ m, is obtainable in 0.85~years.

These results are confirmed by numerical simulation, shown in Fig.~\ref{fig:earth1024}, using an image of the Earth, convolved with the SGL PSF. In the top row, the convolved image was corrupted with Gaussian noise at ${\rm SNR}_{\rm C}=187$ and deconvolved using model parameters $z_0=1.3$~pc, ${\overline z}=1200$~AU. The numerically obtained ${\rm SNR}_{\tt R}=11.4$, slightly better than the estimated value of ${\rm SNR}_{\tt R}=10$, is consistent with our conservative estimate of the deconvolution penalty and the fact that instead of a uniformly illuminated disk, an actual planetary image was used with variable levels of brightness. Similarly, another case depicted in the bottom row with the exo-Earth at 30~pc and imaged at $128\times 128$ pixels with a pre-deconvolution ${\rm SNR}_{\tt C}=41.3$ yields an image with ${\rm SNR}_{\tt R}=3.5$, slightly higher than the estimated value of ${\rm SNR}_{\tt R}=3$.

These examples also highlight why pixel spacing has such a substantial effect on the deconvolution penalty and the resulting integration time. As the middle images in Fig.~\ref{fig:earth1024} show, when pixels are spaced widely apart (i.e., when the convolved image is sampled at a lower resolution), the contribution of blur due to the SGL's spherical aberration is substantially reduced. Even in the convolved image, most features of the Earth's topography are recognizable, only finer details (which would be represented by adjacent pixels) are lost. Consequently, when pixels are far apart, deconvolution introduces a relatively modest penalty, very significantly reducing the required integration time to obtain a high-resolution image of a target exoplanet.

\section{Discussion and Conclusions}
\label{sec:disc}

The SGL is characterized by the geometric properties of the Einstein ring that appears around the Sun, when viewed from a particular vantage point in the SGL's focal region in the vicinity of the primary optical axis at a heliocentric distance $\overline z$, see details in \citep{Turyshev-Toth:2020-extend}. The radius of this Einstein ring (i.e., $b=\sqrt{2r_g \overline z}\geq R_\odot$) and its light collecting area (i.e., for the imaging telescope's aperture $d$, the area ``seen'' by the telescope  is $2\pi b d \geq 2\pi R_\odot d$) determine its angular resolution ($\delta\theta_{\tt SGL}=0.38 \lambda/\sqrt{2r_g \overline z}=1.03\times 10^{-10}(\lambda/1~\mu{\rm m})(650~{\rm AU}/\overline z)^\frac{1}{2}$~arcsec) and light amplification capabilities ($\mu_{\tt SGL}\simeq 2\pi bd/(\pi d^2/4)=8\sqrt{2r_g\overline z}/d=5.57\times 10^9(1~{\rm m}/d)(\overline z/650~{\rm AU})^\frac{1}{2}$), respectively, both of which are many orders of magnitude beyond the capabilities offered by conventional telescopes or even proposed optical interferometric configurations.
For this reason, the SGL is being considered as a means to obtain detailed images of Earth-like exoplanets in other solar systems, at resolutions that cannot be obtained by any other existing or foreseeable technical solution.

A major challenge is that the SGL is not a perfect lens. It is characterized by substantial negative spherical aberration. A further complication is that any light from a distant source appears as an Einstein ring near the solar disk, on the background of the very bright solar corona. However, the SGL's PSF is known, i.e., see \cite{Turyshev-Toth:2020-extend}. This makes it possible, in principle, to reconstruct sharp, detailed images of the observational target. The unwanted background from the solar corona can be measured independently and removed. Inevitably, due to the quantized nature of light, stochastic shot noise with approximately Gaussian characteristics remains. Reducing this noise can only be accomplished by increasing the amount of light collected, using either larger instruments or longer integration times.

In \cite{Turyshev-Toth:2020-extend} we considered an ``annular coronagraph'' concept. Realizing that such a coronagraph cannot be meaningfully implemented in the form of an internal coronagraph, we have not used those results in our present estimates. We do note, however, that the annular occulter concept may be resurrected in the future using an external device, a starshade. For now, we use the most conservative estimates on the contributions of the corona to noise.

We estimated the ``penalty'' incurred by deconvolving images that are blurred due to the spherical aberration of the SGL. This penalty is significant: the nature of the SGL is such that its convolution operator has a high degree of degeneracy, which leads to a disproportionate noise amplification during deconvolution. However, this penalty is significantly mitigated when sampling locations, pixels, in the SGL image plane are farther apart. This is especially the case when considering exoplanet targets in nearby solar systems; the projected image can be up to several ten kilometers in diameter, so even if the image plane is sampled at megapixel resolution, the spacing between image pixels is large, leading to a very significant improvement in the SNR of deconvolved images. Not considering this spacing between image pixels can lead to unrealistically low estimates of deconvolution SNR, as was done by \cite{Willems:2018}.

We have shown that even with the SGL, images obtained of faint, non self-luminous exoplanets are unavoidably photon starved, appearing on top of the bright solar corona. Even if the corona is removed, its contribution to stochastic noise cannot be ignored, as it was done by \cite{MM:2022}; failure to consider this stochastic noise background yields unrealistically high SNR estimates.

We have not considered deviations of the solar gravitational field from the monopole. So long as the resulting caustic pattern is much smaller than the projected image of the target exoplanet (and this is indeed the case in most scenarios under consideration) such deviations are not expected to significantly reduce the pre-deconvolution SNR, since the same amount of light arrives in the image area, just more blurred. The main impact of these multipole contributions is on the deconvolution penalty, a topic that we are studying separately.

We estimated the impact of mission parameters on the resulting integration time. We found that, as expected, the integration time is proportional to the square of the total number of pixels that are being imaged. We also found, however, that the integration time is reduced when pixels are not adjacent, at a rate proportional to the inverse square of the pixel spacing. Consequently, using a fictitious Earth-like planet at the Proxima Centauri system at $z_0=1.3$~pc from the Earth, we found that a total cumulative integration time of less than one year is sufficient to obtain a high quality, megapixel scale deconvolved image of that planet. Furthermore, even for a planet at 30~pc from the Earth, good quality deconvolution at intermediate resolutions is possible using integration times that are comfortably consistent with a realistic space mission with less than one year spent collecting data.

A mission that begins its imaging campaign at 650~AU would benefit from the increasing SNR as it progresses to larger heliocentric distances \citep{Turyshev-etal:2020-PhaseII}. This would allow such a mission to gradually improve its resolution, while keeping integration times sufficiently short to study temporally varying effects, such a diurnal rotation. In addition, the mission would have to be able to compensate for the impact of the solar reflex motion with respect to the solar system barycentric frame. These topics were discussed in \citep{Turyshev-Toth:2022-wobbles}, where we also evaluated the impact of the host star's light contamination on exoplanet imaging. Although the light from the host star is also amplified by the SGL, it is image is formed many tens of kilometers away from that of the target exoplanet, thus allowing for favorable SNR for exoplanet imaging.  The light from the host star occupies two opposing spots on the image sensor of the imaging telescope that may be removed during data processing. As the planet orbits its host star these spots would grow into arclets, then into arcs and then will form the host star's  Einstein ring on top of that from the exoplanet  (see Fig.~16 in \citep{Turyshev-Toth:2022-wobbles}). Clearly, at that time, the imaging operations would have to stop and to be resumed once the planet clears the host star proximity.

There are other interesting challenges. As mentioned, in this study we modeled the solar gravitational field as a monopole field, not yet accounting for deviations in the form of the field's quadrupole and higher moments. We also have not yet accounted for the aforementioned temporal effects, including planetary motion, rotation, varying illumination, light contamination from the host star and interlopers. These issues are rather well-understood and are being  considred.  This work is on-going (e.g., \cite{Turyshev-Toth:2021-multipoles,Turyshev-Toth:2021-caustics,Turyshev-Toth:2021-imaging,Turyshev-Toth:2021-quartic} for multipolar contributions  and \cite{Turyshev-Toth:2022-wobbles} for properly capturing the dynamics), along with work on establishing a technically feasible mission design and architecture. Results will be published elsewhere when available.

\section*{Acknowledgements}

This work in part was performed at the Jet Propulsion Laboratory, California Institute of Technology, under a contract with the National Aeronautics and Space Administration. The authors are
grateful to Karl R. Stapelfeldt and Phil A. Willems of JPL  for motivation and helpful discussions.
VTT acknowledges the generous support of Plamen Vasilev and other Patreon patrons.

\section*{Data Availability}

No data was generated and/or analysed to produce this article.

\bibliographystyle{mnras}

\appendix

\section{Notes on SGL imaging simulations}
\label{appA}

As discussed in the main text, our estimate of the post-deconvolution SNR (\ref{eq:penalty2}) was validated by simulating SGL imaging. For this purpose, we used custom computer code that implemented Fourier convolution and deconvolution using the SGL PSF.

The advantage of Fourier-deconvolution is that it is accomplished by simple division in Fourier space; in image space, it would require the inversion of an $N\times N$ matrix (or solving a linear system of $N$ equations in $N$ variables) where $N$ is the total number of pixels (i.e., for a $1024\times 1024$ pixel image, $N=1024\times 1024=1048576$.) Direct deconvolution of even a $128\times 128$-pixel image (the largest that can be practically deconvolved on desktop-class hardware) requires many hours; larger images would require dedicated hardware or supercomputing facilities.

Fourier-deconvolution can, however, introduce artifacts. In the case of our simulation this is mitigated by the fact that we use the same Fourier method for convolution, and the artifacts cancel. Thanks to this, we found that Fourier methods can be used reliably to analyze the SNR costs (``penalty'') associated with deconvolution, making it possible to economically investigate a large number of cases. We have validated the method at lower resolutions by directly comparing Fourier and direct convolution and deconvolution, further confirming that estimates of SNR, obtained using Fourier methods, can be trusted.

\begin{figure}
\centering\includegraphics{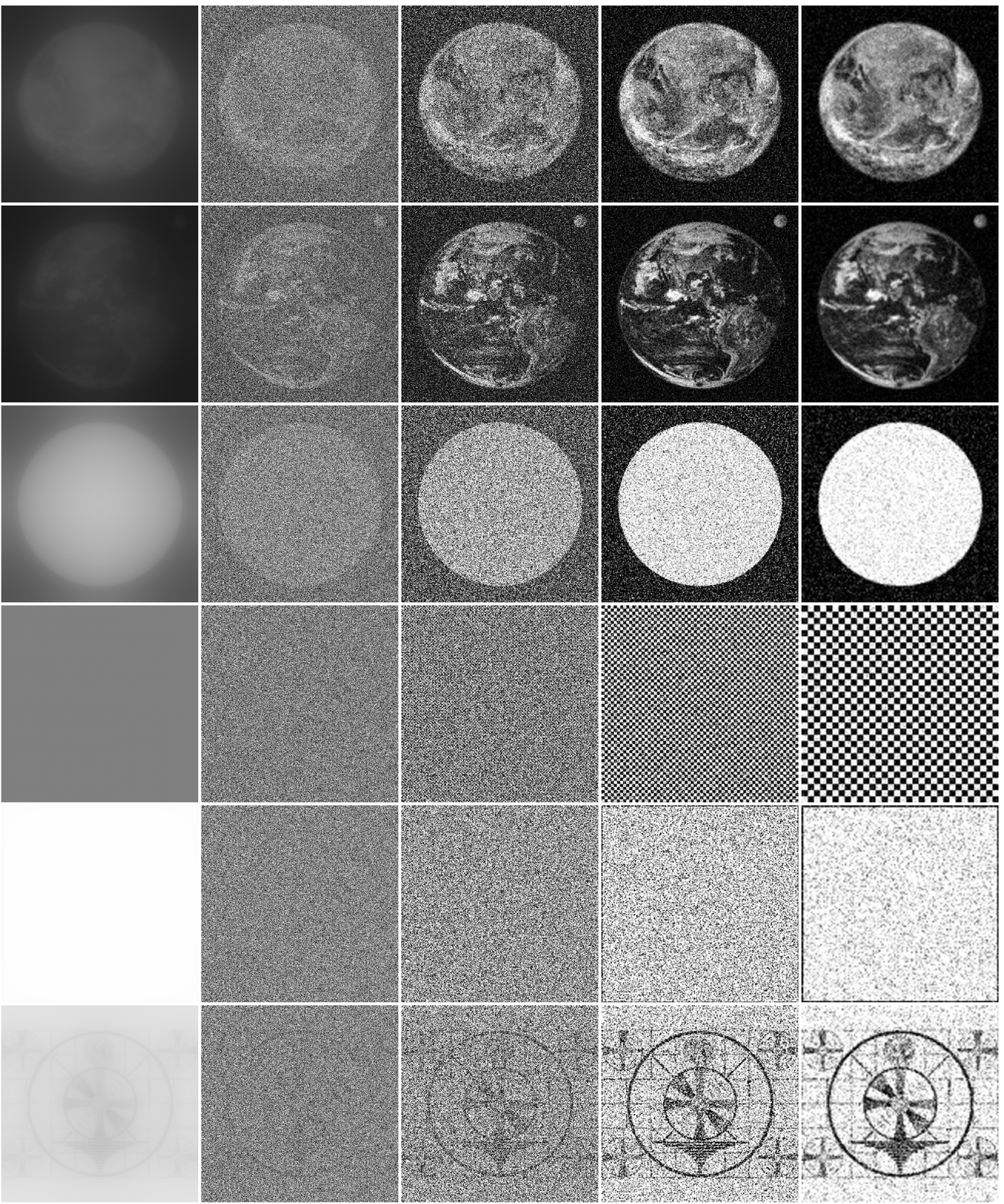}\par
\caption{\label{fig:images-m}A representative sample of images from a simulation run involving 7 patterns, 5 different projected image sizes, 4 resolutions and 9 different pre-deconvolution SNR levels. The rows correspond to a monochrome image of the Earth, a high-contrast image of the Earth-Moon system, a uniform white disk, a checkered pattern, an all-white background and a classic television test pattern. Columns contain the convolved image followed by deconvolution results at resolutions of $1024\times 1024$, $512\times 512$, $256\times 256$ and $128\times 128$ pixels.}
\end{figure}

To study the monopole lens, we created a number of test cases using both images of the Earth (acting as a ``stand-in'' for a would-be exoplanet) and other test patterns. We ran a large number of cases to check the validity and robustness of our SNR estimates. Images in the present paper were produced using a test run that involved the following:
\begin{itemize}
\item Seven test images:
  \begin{inparaenum}[1)]
  \item the fully illuminated Earth;
  \item a high-contrast image of the Earth-Moon system;
  \item a uniformly illuminated disk;
  \item a checkered test pattern;
  \item an all-white pattern;
  \item an all-black pattern;
  \item a classic television test pattern.
  \end{inparaenum}
\item Five different image sizes (diameters):
  \begin{inparaenum}[1)]
  \item 1024 meters, corresponding to tightly packed image pixels when a $d=1$~m aperture is used for observation;
  \item 1300 meters, corresponding to an exo-Earth at $\sim$30~pc with the SGL telescope at 650~AU;
  \item 10,000 meters, corresponding to an exo-Earth at $\sim$4~pc with the SGL telescope at 650~AU;
  \item 43,000 meters, corresponding to an exo-Earth at $\sim$1.3~pc, with the SGL telescope at 900~AU; and
  \item 57,000 meters, corresponding to an exo-Earth at $\sim$1.3~pc, with the SGL telescope at 1200~AU.
  \end{inparaenum}
\item Nine different levels of Gaussian noise added prior to deconvolution (simulating light contamination affecting the observation of the convolved image): $SNR=10$, 30, 100, 300, 1000, 3000, 10000, 30000 and 100000.
\item Four different image resolutions: $128\times 128$, $256\times 256$, $512\times 512$ and $1024\times 1024$ pixels (powers-of-two were used for compatibility with the fast Fourier transform.)
\end{itemize}
The deconvolution result from representative sample of these simulations is shown in Fig.~\ref{fig:images-m}.

Across all results, we found that the deconvolution penalty remained consistent. In retrospect, this is hardly surprising: both convolution and deconvolution are linear processes (modeled by multiplying the signal pixels by the convolution matrix or its inverse) thus the effect of deconvolution on the signal are independent from the effects of deconvolution on the added noise.

Comparing the SNR of deconvolved images against the values predicted by Eq.~(\ref{eq:penalty2}) we observe the following:
\begin{itemize}
\item When pixels are adjacent or nearly so, Eq.~(\ref{eq:penalty2}) accurately predicts the penalty: simulations consistently yield a penalty of 0.877, very close to the predicted value of 0.891.
\item When pixel spacing is increased, simulations consistently yield slightly better SNRs than the value predicted by Eq.~(\ref{eq:penalty}).
\item When pixel spacing is so large that Eq~(\ref{eq:penalty2}) would predict unrealistic ``penalty'' (i.e., an actual improvement in SNR as a result of deconvolution, which of course will not happen) simulated results plateau around the achievable maximum (corresponding to identical pre- and post-deconvolution SNRs.)
\end{itemize}
Therefore, we consider this behavior well-understood, confirming that Eq.~(\ref{eq:penalty2}) indeed yields a valuable, reliable estimate of the post-deconvolution SNR despite the aggressively simplified model of the convolution matrix that was used in the development of this expression.

For more details and a complete set of simulation results, see \url{https://sglf.space/MONOSGL/results.html}.

\label{lastpage}

\end{document}